\documentclass[a4paper,twocolumn,11pt, accepted=2026-10-26]{quantumarticle}

\pdfoutput=1
\usepackage{graphicx} 
\usepackage{physics}
\usepackage{xcolor}
\usepackage{amsmath} 
\usepackage{amssymb} 
\usepackage{float}
\usepackage{svg}
\usepackage{bm}
\usepackage{setspace}
\usepackage{hyperref}
\usepackage{cleveref}

\usepackage{algorithm,algorithmic}

\usepackage[percent]{overpic}

\begin{document}
\title{Optimal Calibration of Qubit Detuning and Crosstalk}

\author{David Shnaiderov}
\affiliation{Department of Physics, Bar-Ilan University, 52900 Ramat Gan, Israel}

\author{Matan {Ben Dov}} 
\affiliation{Department of Physics, Bar-Ilan University, 52900 Ramat Gan, Israel}

\author{Yoav Woldiger}
\affiliation{Department of Physics, Bar-Ilan University, 52900 Ramat Gan, Israel}
\affiliation{Institute of Nanotechnology and Advanced Materials,  Bar-Ilan University, 52900 Ramat Gan, Israel}

\author{Assaf {Hamo}} 
\affiliation{Department of Physics, Bar-Ilan University, 52900 Ramat Gan, Israel}
\affiliation{Institute of Nanotechnology and Advanced Materials,  Bar-Ilan University, 52900 Ramat Gan, Israel}

\author{Eugene {Demler}}
\affiliation{Institute for Theoretical Physics, ETH Z\"urich, 8093 Z\"urich, Switzerland}

\author{Emanuele G. Dalla Torre}
\affiliation{Department of Physics, Bar-Ilan University, 52900 Ramat Gan, Israel}


\begin{abstract}

Characterizing and calibrating physical qubits is essential for maintaining the performance of quantum processors. A key challenge in this process is crosstalk, which complicates the estimation of individual qubit detunings. In this work, we derive optimal strategies for estimating detuning and crosstalk parameters by optimizing Ramsey interference experiments using Fisher information and the Cramér–Rao bound. We compare several calibration protocols, including measurements of a single quadrature at multiple times and of two quadratures at a single time, for a fixed number of total measurements. Our results predict that the latter approach yields the highest precision and robustness in both cases of isolated and coupled qubits. We validate our approach experimentally using both a single NV center and superconducting transmon qubits. Our approach enables accurate parameter extraction with significantly fewer measurements, resulting in up to a 50\% reduction in calibration time while maintaining estimation accuracy.
\end{abstract}

\maketitle

\section{Introduction}

{\color{black}
Operating a quantum processor at high performance requires accurate tuning of its control parameters. In realistic devices, qubit transition frequencies, coherence properties, and effective couplings drift because of fluctuations in materials, the electromagnetic environment, and control hardware \cite{kjaergaard2020superconducting,siddiqi2021engineering,chatterjee2021semiconductor}. As systems scale, calibration therefore becomes a recurring task rather than a static characterization step, motivating automated, feedback-based, and low-latency recalibration workflows across quantum platforms \cite{wittler2021integrated,pasquale2024qibocal,marciniak2026millisecond}.

Among the parameters that must be monitored, qubit frequencies play a particularly direct role because they introduce coherence errors even during idle evolution: a mismatch between the qubit transition frequency and the control frame leads to unwanted phase accumulation, which appears as an effective Z rotation and must be repeatedly estimated and compensated. This makes qubit-detuning calibration a useful representative example of the broader recalibration problem. Recent work has shown that frequency estimation can benefit from adaptive Ramsey-style experiment design and real-time feedback \cite{mcmichael2021sequential,berritta2025efficient}, and that frequency drift can require recalibration on very short time scales \cite{marciniak2026millisecond}. These results highlight the measurement efficiency of the underlying Ramsey-type estimation loop as an important resource.

In multiqubit superconducting processors, detuning calibration is further complicated by residual crosstalk \cite{gicev2026crosstalk}. Static ZZ interactions and control-line crosstalk shift the effective frequency of one qubit depending on the state of neighboring qubits \cite{mundada2019suppression,kosen2024signal,fors2024comprehensiveexplanationzzcoupling}. As a result, detuning calibration can no longer be reduced to independent simultaneous single-qubit measurements: the relevant quantities are conditional frequency shifts that must be isolated by an appropriate measurement schedule. This challenge has motivated a broad literature on crosstalk detection and characterization \cite{sarovar2020detecting,rudinger2021experimental}, scalable calibration of flux and signal crosstalk \cite{barrett2023learning,kosen2024signal,fan2025calibrating}, physical understanding of residual ZZ coupling \cite{fors2024comprehensiveexplanationzzcoupling}, and hardware or pulse level suppression strategies \cite{mundada2019suppression,ni2022scalable,xie2022suppressing,alghadeer2026low}.


In this work, we optimize this shared estimation task. We consider Ramsey-type calibration experiments for estimating single-qubit detunings and crosstalk-induced conditional frequency shifts. Using Fisher information and the Cramér-Rao bound, we identify measurement times and quadratures that maximize the information obtainable within a fixed measurement budget. This statistical-design perspective is well-suited to recurring calibration, where the same estimation problem must be solved repeatedly, and each unnecessary measurement directly contributes to calibration overhead.

We first analyze the single-qubit detuning problem and show how the optimal choice of measurement time depends on the unknown detuning, dephasing, and prior information available from previous calibration rounds. We then extend the framework to multiqubit settings with residual ZZ type interactions, in which the detuning of a target qubit depends on the computational state of its neighbors. In this case, the measurement schedule must resolve several conditional frequencies rather than a single offset. We show how Fisher information provides a way to choose the input configurations, evolution times, and measurement quadratures that best distinguish these parameters under a finite shot budget.

The resulting protocol can be incorporated directly into existing automated calibration, crosstalk mitigation, and software compensation pipelines \cite{kakkar2025no,pack2026benchmarking}. Rather than replacing these approaches, it reduces the experimental overhead of the parameter-estimation step that they depend on. In this sense, optimal experimental design provides a practical route to faster, more precise recurring calibration of quantum processors subject to drift and residual crosstalk.

}

\section{Single qubit model}

To introduce our optimal strategy, we first consider the case of a single qubit, whose detuning $\omega$ is unknown. To mimic realistic conditions, we assume that the qubit undergoes a dephasing process. The dynamics of the qubit is then described by 
\begin{align}
	H = \frac{\omega}{2}(1-Z) + \frac{h(t)}{2}(1-Z). \label{eq:H1q}
\end{align}
Here, \(Z\) is a Pauli matrix 
with eigenvalues $+1$ and $-1$, respectively for the $|0\rangle$ and $|1\rangle$ states,
and \(h(t)\) is a Gaussian random process with $\langle h(t)\rangle=0$ and two-point correlation function $\langle h(t)h(t')\rangle = F(t-t')$
\cite{breuer2002theory,benedetti2014effective}. 

The common procedure to calibrate the qubit consists of a series of Ramsey interference experiments, where (i) an initial $\pi/2$ pulse prepares the qubit in the superposition state, $|+\rangle = (|0\rangle + |1\rangle)/\sqrt{2}$; (ii) the qubit is let evolve freely for time $t$; and (iii) the Pauli operator $X$ is measured by applying a second $\pi/2$ pulse and measuring the qubit in the computational basis. The experiment is repeated for varying $t$, and the average result is stored as $\langle X(t)\rangle_{\rm exp}$. This quantity is then compared to the theoretical result obtained by evolving the initial state with Eq.~(\ref{eq:H1q}), $|\psi\rangle = \left[|0\rangle + \exp(-i\omega t-i\int_0^t dt'~ h(t'))|1\rangle\right]/\sqrt{2}$ and averaging over $h(t)$, {\color{black}leading to 
$\langle X(t)\rangle_{\rm theory} = \cos(\omega t) \exp\left(-\tfrac{1}{2}\int_0^t dt' \int_0^t dt'' F(t'-t'')\right)$. If the noise correlations have short memory, one can approximate $F(t-t')=2\gamma\delta(t-t')$, where $\gamma$ is the dephasing rate and $\delta(t-t')$ is the Dirac delta. In this case, often referred to as the Markovian limit, one obtains a closed expression that depends only on $\omega$ and $\gamma$:
\begin{align}
\langle X(t)\rangle_{\rm theory} = \cos(\omega t)e^{-\gamma t}
\label{eq:Xtheory}
\end{align}
These two parameters are then estimated by minimizing the difference between $\langle X(t)\rangle_{\rm exp}$ and $\langle X(t)\rangle_{\rm theory}$ \cite{qiskit_calibration}.

{\color{black}
If the quantum bath has a memory time comparable to $1/\gamma$, the Markovian approximation is no longer valid, and the Ramsey decay cannot be described by a single exponential envelope. In
that regime, the relevant noise correlations, or equivalently, the noise
spectrum, must first be characterized, for example, using quantum noise
spectroscopy \cite{clerk2010introduction}. The Ramsey signal should then be compared with the corresponding
non-Markovian response function,
\begin{align}
\langle X(t)\rangle_{\rm theory}=f(t;\boldsymbol{\theta}),
\end{align}
where \(\boldsymbol{\theta}\) may include \(\omega\), \(\gamma\),
\(\tau_{\rm bath}\), or other parameters describing the bath. In this sense,
finite-memory noise can be incorporated into our optimization procedure by
replacing the Markovian expression in Eq.~(\ref{eq:Xtheory}) with the
appropriate Ramsey response function, rather than by assuming an exponential
decay.

}
\begin{figure}[t]
    \centering
    \includegraphics[width=\linewidth]{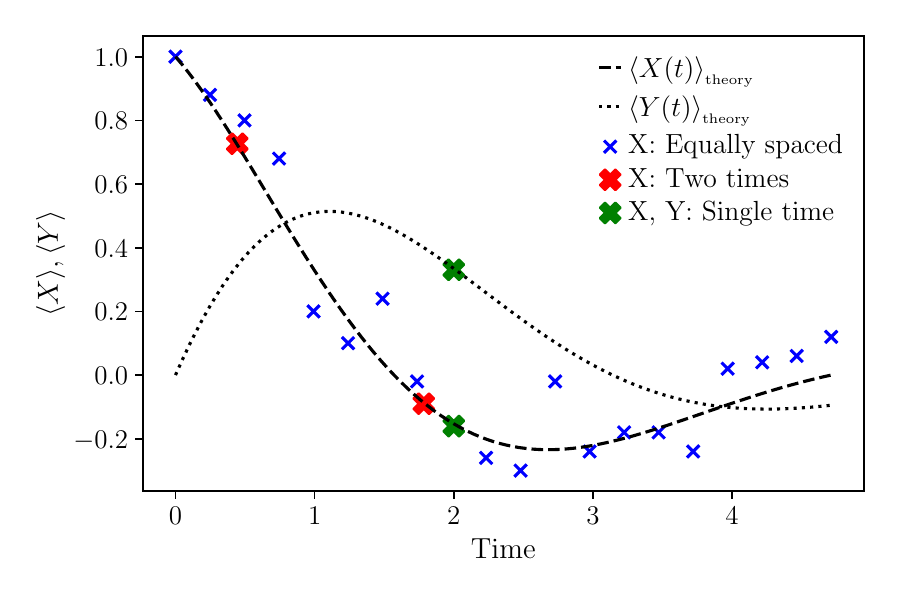}
    \vspace{-0.5cm} 
    \caption{Four strategies to calibrate the frequency $\omega$ and dephasing rate $\gamma$ of a single qubit, respectively denoted by blue, purple, red, and green crosses, see text for details. The blue and purple strategies are noisier because they use more measurement times and fewer shots.}
    \label{fig:schematic}
\end{figure}


%
%


In a real experiment, the precision of the above-mentioned procedure is limited by {\color{black} different sources of noise, such as state preparation and measurement (SPAM) errors, imperfections in the $\pi/2$ pulses, and shot noise. The former two types of noise do not depend on the number of measurement times, $N_{\rm times}$ and, for state-of-the-art quantum computers with single-qubit gate fidelity of $99.9\%$, limit the calibration precision to about $10^{-3}$ or less. Shot noise refers to the fact that each individual measurement (or, {\it shot}) of $X(t)$ returns $+1$ and $-1$ and leads to a standard deviation that scales with $1/\sqrt{N_{\rm shots}}$}. For a given total number of shots $N_{\rm tot}=N_{\rm times}N_{\rm shots}$, one should allocate this resource wisely among the different measurement times. A fundamental question that we address is whether one obtains a better precision by performing many measurements at fewer times (see red crosses in Fig.~\ref{fig:schematic} for $N_{\rm times}=2$ and $N_{\rm shots}=1000$), or on the contrary, by spreading the measurements over a larger number of times (see blue crosses in Fig.~\ref{fig:schematic} for $N_{\rm times}=20$ and $N_{\rm shots}=100$).

\section{Methods: Cramér-Rao bound}

Motivated by earlier studies on the optimization of NMR experiments \cite{JONES199625}, we address this question using the Fisher information \cite{ly2017tutorialfisherinformation}, $\mathcal{I}$, which quantifies how much information a set of measurements provides about an unknown parameter of interest. For a set of variables $\bm{X}$ whose probability distribution $f(\bm{X}|\bm{\theta})$ depends on a set of parameters $\bm\theta$, $\mathcal{I}$ is defined as
\begin{align}
	\mathcal{I}_{jk}
    ({\mathbf{\theta}}) = \mathbb{E}_{\bm X}\left[\frac{\partial}{\partial \theta_j} \ln L(\bm \theta|\bm X)\frac{\partial}{\partial \theta_k} \ln L(\bm\theta|\bm X)\right], 
\end{align}
where \(L(\bm\theta|\bm X) = f(\bm X|\bm \theta)\) is the likelihood function and $\mathbb{E}_{\bm X}$ is the weighted average over $\bm X$.
The Cramér-Rao bound states that the covariance of an unbiased  estimator $\hat{\bm\theta}$ is bounded from below by the inverse of the Fisher information matrix,
\begin{align}
    {\rm cov}(\hat{\bm\theta}) \geq \mathcal{I}^{-1}.
    \label{eq:CRB}
\end{align}

In the problem at hand, the unknown parameters to be estimated from the data are 
$\bm{\theta} = (\omega,\gamma)$, and the measurement outcomes are 
$\mathbf{X}=(X_1, X_2, ..., X_{N_{\rm tot}})$, where $X_i$ is a single, independent, 
binary measurement outcome at time $t_i$ with mean value $\langle X(t_i) \rangle$, 
given in Eq.~(\ref{eq:Xtheory}). The measurement times $\{t_i\}$ are design variables, they are chosen by the experimenter prior to the measurement, 
and each time is probed with a single shot. Repeated times are allowed, so that 
allocating $n_i$ shots to a time $t_i$ is equivalent to choosing $t_i$ to appear 
$n_i$ times in the list $\{t_1, \ldots, t_{N_{\rm tot}}\}$. The Fisher information 
matrix therefore depends on both the unknown parameters and the chosen times,
\begin{align}
    \mathcal{I}_{\alpha\beta}(\omega,\gamma;\{t_i\}), \qquad \alpha,\beta \in \{\omega,\gamma\},
\end{align}
and we optimize the calibration strategy by {\it minimizing} the Cram\'er-Rao bound 
over the design variables $\{t_i\}$, i.e. by minimizing 
${\rm Tr}[\mathcal{I}^{-1}(\omega,\gamma;\{t_i\})]$ with respect to $\{t_i\}$, 
for fixed $\bm{\theta}=(\omega,\gamma)$.




Our analysis is conceptually related to previous work on optimal measurement strategies and frequency estimation in passive atomic clocks \cite{rosenbandExponentialScalingClock2013,mullanOptimizingPassiveQuantum2014,fraasAnalysisStationaryOperation2016a,lerouxOnlineEstimationLocal2017,kielinskiBayesianFrequencyMetrology2025}. In that setting, Ramsey sequences and feedback servos are optimized to counter stochastic fluctuations of a local oscillator (LO) relative to a stable atomic transition, with performance quantified in terms of long-term clock stability.

The present work differs from these earlier studies in three key aspects: First, we do not assume an independently steerable LO or an online servo loop, but rather optimize the choice of measurement times and measured quadratures of each qubit separately.
Second, passive clocks are generically analyzed in the presence of temporally
correlated local-oscillator noise. In the present work, by contrast, we focus on
the standard exponential Ramsey-decay model as an effective
description of dephasing over the calibration timescale of interest. This
corresponds to extracting an effective \(T_2^*=1/\gamma\) from a decaying
Ramsey oscillation, and is supported by the experimental Ramsey traces analyzed
below, which are well described by decaying-cosine fits. 

In this limit, earlier studies are no longer applicable and predict divergent optimal interrogation times \cite{fraasAnalysisStationaryOperation2016a}.
Thanks to these two simplifications, we obtain analytical expressions for the optimal measurement times without the need for numerical elaboration of real-time feedback from the quantum device. Finally, our protocols make explicit use of the multi-qubit structure and conditional frequency shifts that do not have a direct analog in single-ensemble clock architectures and are designed to be directly deployable within quantum computing calibration pipelines.

{\color{black} We emphasize, however,
that this assumption is not universal: superconducting qubits and solid-state
spin qubits may also exhibit correlated dephasing noise, including
\(1/f\)-type noise, quasiparticle-induced fluctuations, and spin-bath-induced
noise \cite{ithier2005decoherence, bylander2011noise, serniak2018hot, sung2019non, chan2020twoqubit}. In such cases, the simple exponential Ramsey envelope in
Eq.~(\ref{eq:Xtheory}) should be replaced by the appropriate non-Markovian
response function, which may be obtained from a microscopic model or from
quantum noise spectroscopy. The optimization used below can
then be applied to that modified response function.}

\section{Optimal measurement times}
The first question that we address is the optimal number of {\it distinct} measurement times. To study this problem, we start from a set of randomly distributed $t_i$, each measured $n_i=1$ times, and optimize them numerically by minimizing the CRB, Eq.(\ref{eq:CRB}). After the optimization, any two times $t_i$ and $t_j$ that approach each other within numerical precision are grouped and combined into a single measurement time measured $n_i+n_j$ times, see Algorithm~\ref{alg:optimization}. This protocol conserves the total number of measurements while reducing the number of measurement times. When applied to Eq.~(\ref{eq:Xtheory}), this approach converges to only two measurement times.

\begin{algorithm}[H]
\caption{Optimal Allocation of Measurement Times and Shots}
  \label{alg:optimization}
  \begin{spacing}{1.2}
  \begin{algorithmic}[1]

    \REQUIRE Maximal number of measurement times $N_{\rm times}$ \hfill (default $N_T=10$)
    \REQUIRE Merge tolerance $\delta$                              \hfill (default $\delta = 0.01$)
    \ENSURE  Optimised times and shots $(\widehat T,\widehat N)$

    \STATE \textbf{Initialise:} sample $N_{\rm times}$ times
    $\{t_i\}$ randomly and set $n_i = 1$ for all $i$

    \STATE \textbf{Optimise:} \hspace{\algorithmicindent}%
      $\displaystyle
        (\widehat T,\widehat N)=$ \\
        $\arg\min_{\{t_i,n_i\}}
          \bigl[\,\mathrm{CRB}_{\omega}(T,N)
                +\,\mathrm{CRB}_{\gamma}(T,N)\bigr]$%
      \\
      \textbf{s.t. }%
      $\sum_i n_i = N_{\rm T},\;
        n_i \in \mathbb{N}_{\ge 0},\;
        t_i>0$

    \STATE \textbf{Group:} merge any $t_i,~ t_j$ with $|t_i-t_j|<\delta$\\
      by
      $\;(t_i,t_j)\!\longrightarrow\!\bigl((t_i+t_j)/2\bigr)$  and \\
      $(n_i,n_j)\!\longrightarrow\!n_i+n_j$

    \RETURN $(\widehat T,\widehat N)$

  \end{algorithmic}
  \end{spacing}
  
\end{algorithm}

The optimal value of the Fisher information bounds from below by the precision of the calibration process through the CRB, Eq.~(\ref{eq:CRB}). Figure \ref{fig:Theoretical_bound}(a) shows this bound as a function of $t_1$ and $t_2$ for $\omega=\gamma$. This function is a single minimum at $t_1 \approx 0.44/\gamma$ and $t_2 \approx 1.78/\gamma$, where the calibration is optimal. The minimum around this point is very broad, indicating that the calibration procedure is fairly insensitive to the precise values of $t_1$ and $t_2$. To investigate the dependence of the optimal times on the system's parameter, we repeat the same procedure for varying $\omega/\gamma$ and show the resulting optimal times in the inset of Fig.~\ref{fig:ratios}.

\begin{figure}[t]
	\centering
	\begin{tabular}{l} 
    \begin{overpic}[width=0.95\linewidth]{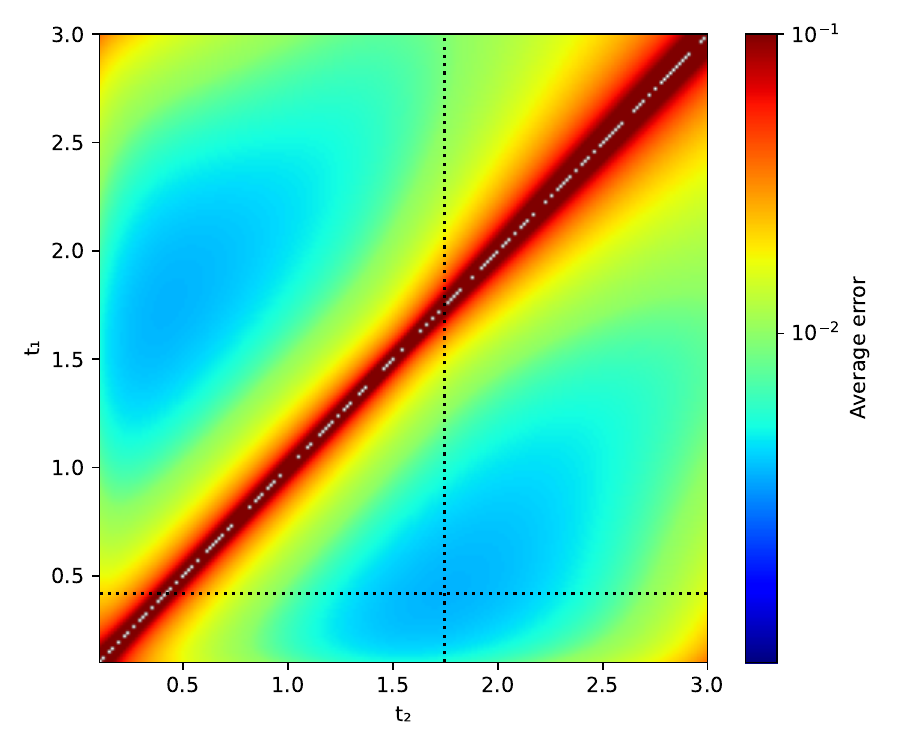}
    \put(0,80){(a)}
  \end{overpic}\\
    \vspace{-0.5cm}
    \begin{overpic}[width=0.95\linewidth]{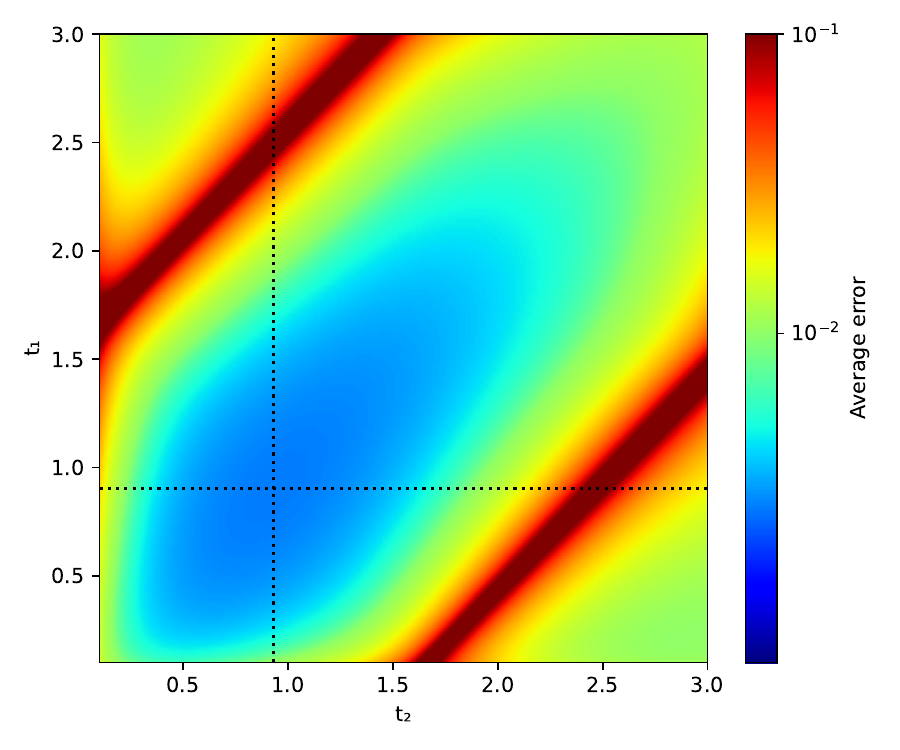}
    \put(0,80){(b)}
   \end{overpic}
   \end{tabular}
    \vspace{-0.4cm}
	\caption{
    Cramer--Rao bound (CRB) for the joint estimation of $\omega$ and $\gamma$ as a function of two measurement times $t_1$ and $t_2$. (a) Both measurements are performed in the $X$ basis, $X(t_1)$ and $X(t_2)$. (b) Orthogonal-quadrature measurements, $X(t_1)$ and $Y(t_2)$. Access to both quadratures reduces parameter correlations and localizes the optimal measurement strategy, with the minimum occurring at two nearby times clustered around the dephasing timescale $t\sim 1/\gamma$.
    }
    \label{fig:Theoretical_bound}
\end{figure}

The optimal strategy consists of a different number of measurements (shots) for the two measurement times, denoted by $n_1$ and $n_2$. In the main panel of Fig.~\ref{fig:ratios} we plot their ratio, as a function of $\gamma/\omega$, varying from $n_1/n_2\approx 1$ for $\gamma\lesssim\omega$, to $n_1/n_2\approx 0.6$ for $\gamma=3\omega$. Intuitively, because $t_2$ is more affected by the dephasing process than $t_1$, one needs to invest more shots to extract useful information at later times, and $n_2>n_1$.
 
To further improve the calibration process, one can measure different quadratures at different times. See also \cite{zohar2023real,arshad2024real} for an adaptive protocol to optimize the measured quadrature based on previous measurements in real-time experiments. In quantum computers, real-time controls are challenging, and we instead consider a simple strategy in which the $X$ and $Y$ quadratures are measured respectively at times $t_X$ and $t_Y$. Our optimization protocol, Algorithm \ref{alg:optimization}, delivers Fig.~\ref{fig:Theoretical_bound}(b), which has a single, broad minimum at $t_X \approx 0.90/\gamma$ and $t_Y\approx 0.93/\gamma$.

\begin{figure}[t]
    \centering
    \includegraphics[width=0.95\linewidth]{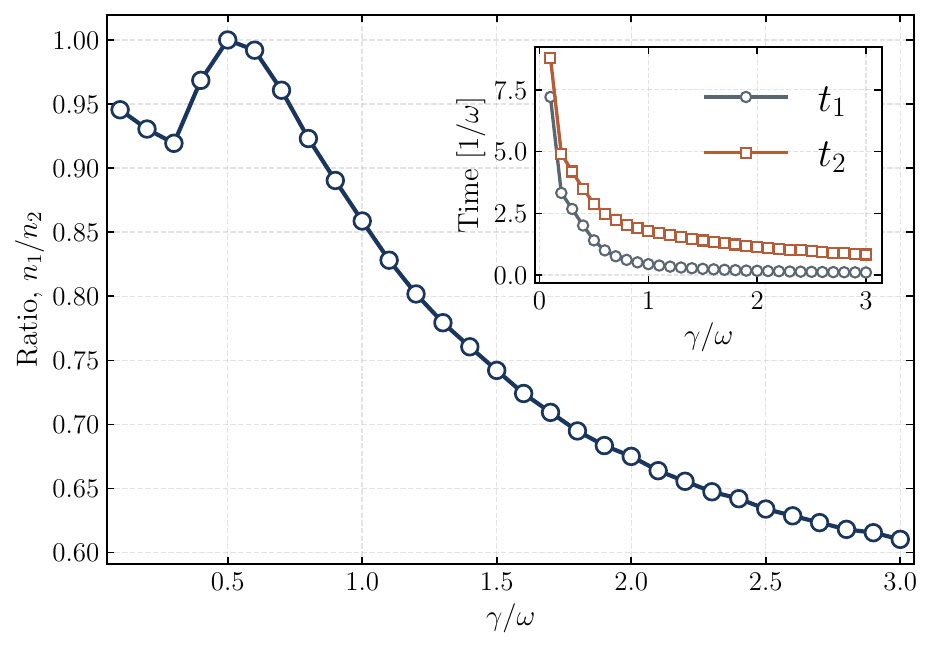}
    \caption{Numerically optimized strategy for the calibration of a qubit with decay $\gamma$ and detuning $\omega$, using $X$ measurements at two times. The main graph shows the optimal ratio of the number of measurements, $n_1/n_2$, as a function of $\gamma/\omega$. The inset shows the corresponding optimal measurement times $t_1$ and $t_2$.}
    \label{fig:ratios}
\end{figure}

This numerical result can be justified analytically using a self-consistent Gaussian approximation valid in the limit of a large number of measurements: 
According to the central-limit theorem, the average over a large number of $Q=\pm1$ measurements is well described by a Gaussian distribution with mean $\expval{Q(t)}$ and variance ${\rm Var}[Q]=(1 - \expval{Q(t)}^2)/{N_{Q}}$ where $N_{Q}$ is the number of shots performed in the $Q$ basis. Under this approximation, the Fisher information is given by \cite{JONES199625}:
\begin{align}
    \mathcal{I}^{(Q)}_{ij}(t)
    = \frac{\partial_{\theta_i}\langle Q(t)\rangle
    \partial_{\theta_j}\langle Q(t)\rangle}
    {{\rm Var}[Q]}.
    \label{eq:FI_qubit_single}
\end{align}

In our case $\theta=(\omega,\gamma)$ and
\begin{equation}
\begin{aligned}
\mathcal{I}_X(t)
&=
\frac{1}{{\rm Var}[X]}
\begin{pmatrix}
(\partial_\omega\langle X\rangle)^2 & (\partial_\omega\langle X\rangle)(\partial_\gamma\langle X\rangle)\\
(\partial_\gamma\langle X\rangle)(\partial_\omega\langle X\rangle) & (\partial_\gamma\langle X\rangle)^2
\end{pmatrix} \\
\mathcal{I}_Y(t)
&=
\frac{1}{{\rm Var}[Y]}
\begin{pmatrix}
(\partial_\omega\langle Y\rangle)^2 & (\partial_\omega\langle Y\rangle)(\partial_\gamma\langle Y\rangle)\\
(\partial_\gamma\langle Y\rangle)(\partial_\omega\langle Y\rangle) & (\partial_\gamma\langle Y\rangle)^2
\end{pmatrix}.
\end{aligned}
\end{equation}

If we self-consistently assume that $t\approx1/\gamma$, we have $e^{-2\gamma t} \approx 0.135 \ll 1$, allowing us to approximate ${\rm Var}[X] = (1-\cos^2(\omega t)e^{-2\gamma t})/N_X\approx 1/N_X$ and similarly for ${\rm Var[Y]}\approx 1/N_Y$. Using
\begin{equation}
\begin{aligned}
\partial_{\omega}\langle X(t)\rangle &= \partial_{\gamma}\langle Y(t)\rangle = -t e^{-\gamma t}\sin(\omega t), \\
\partial_{\gamma}\langle X(t)\rangle &= -\partial_{\omega}\langle Y(t)\rangle = -t e^{-\gamma t}\cos(\omega t), \\
\end{aligned}
\end{equation}
we obtain that for $N_X=N_Y$ the total Fisher information matrix simplifies to a unit matrix
\begin{align}
\mathcal{I}(t) \equiv \mathcal{I}_X(t) +\mathcal{I}_Y(t) \approx N_Q t^{2}e^{-2\gamma t}\,\mathbb{I}_{2\times 2},
\end{align}
This function is maximal precisely at $t^{\ast}=1/\gamma$, in agreement with our numerical result.





\begin{figure}[t]
	\centering
	\begin{tabular}{c}		            
    \hspace{-0.5cm}\begin{overpic}[width=1\linewidth,trim={0, 0.2cm, 0, 0}]{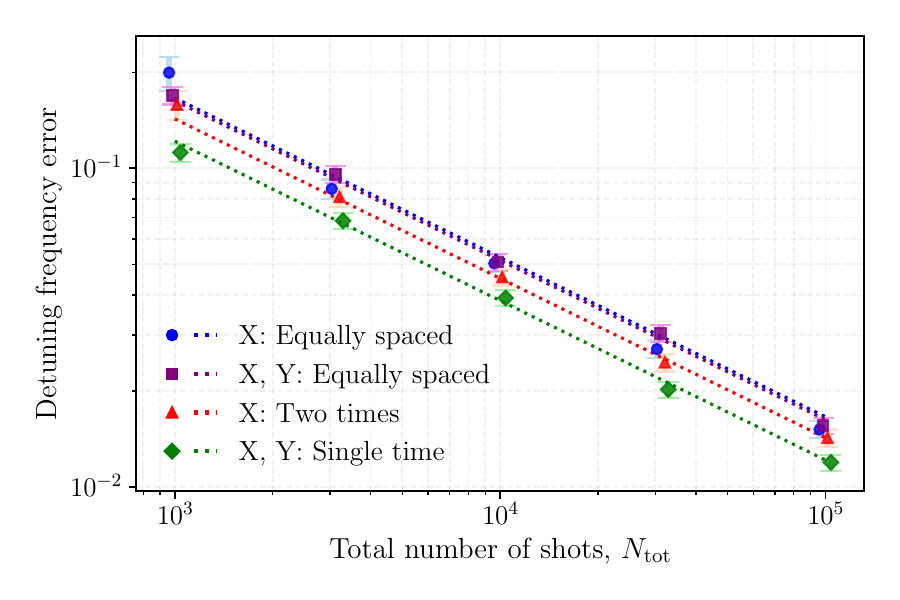}
    \put(5,60){(a)}
    \end{overpic}\\

    \begin{overpic}[width=\linewidth,trim={0, 0.5cm, 0, 0.5cm}]{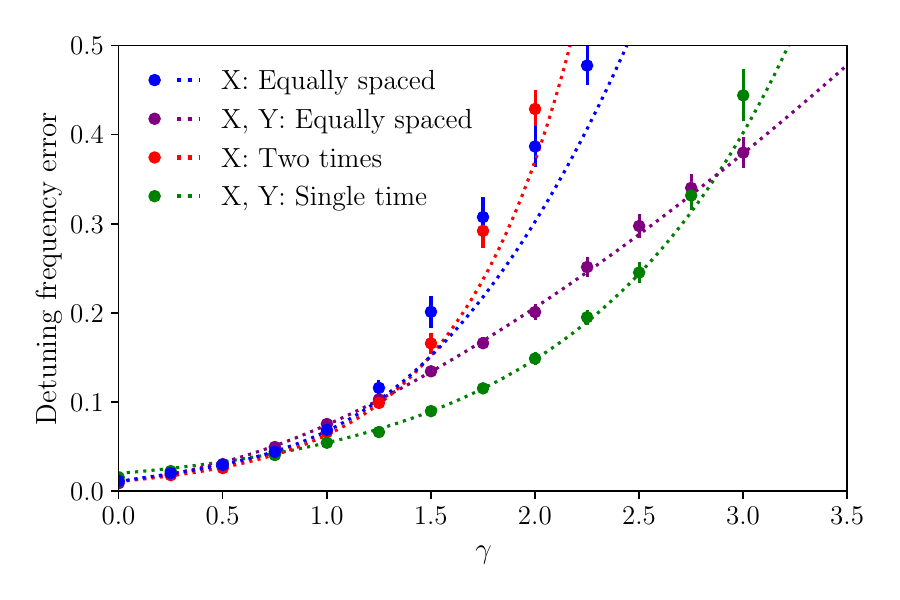}
    \put(0,60){(b)}
    \end{overpic}
    \vspace{-0.2cm}
	\end{tabular}
	\caption{Numerical simulation of four strategies for estimating the qubit detuning and dephasing rate as a function of (a) the total number of shots, $N_{\rm tot}=N_{\rm times}N_{\rm shots}$, (b) actual dephasing rate (see text for details). The circles are numerical results, and the dotted lines are the Cramér-Rao bound. The optimal strategy (green) involves measuring both $X$ and $Y$ at $t=1/\gamma$.}

    \label{fig:single_qubit_noise_scaling}
\end{figure}

\section{Single qubit simulation}


{\color{black}
To validate our optimization procedure, we compare the theoretical bound with numerical simulations of the Markovian-dephasing model. In this limit, the stochastic Hamiltonian in Eq.~(\ref{eq:H1q}) is replaced, after noise averaging, by an equivalent Lindblad master equation for pure dephasing,
\begin{align}
\dot{\rho}=-i\left[\frac{\omega}{2}(1-Z),\rho\right]+\frac{\gamma}{2}\left(Z\rho Z-\rho\right),\label{eq:lindblad_single_qubit}
\end{align}
which corresponds to the Lindblad jump operator\begin{align}L=\sqrt{\frac{\gamma}{2}}\,Z .\end{align}The dissipative term suppresses the off-diagonal density-matrix elements as\begin{align}\rho_{01}(t)\propto e^{-\gamma t},\end{align}while the Hamiltonian term produces the coherent phase rotation at frequency \(\omega\). Therefore, for the initial state \(|+\rangle\), the master equation gives\begin{align}\langle X(t)\rangle=\cos(\omega t)e^{-\gamma t},\end{align}consistent with Eq.~(\ref{eq:Xtheory}). We solve this master equation using the QuTiP Python library \cite{johansson2012qutip} and introduce shot noise by sampling binary measurement outcomes from the resulting density matrix \cite{grynberg2010introduction}. We use the noisy numerical result to compute \(\langle X(t)\rangle_{\rm noisy}\) and fit it to Eq.~(\ref{eq:Xtheory}) by minimizing the mean-square error between the two curves. The extracted \(\gamma_{\rm noisy}\) and \(\omega_{\rm noisy}\) are then compared to the input values.
}


Typical results from this approach are shown in Fig.~\ref{fig:single_qubit_noise_scaling} (a), where we chose $\omega=\gamma=1$. This figure shows the relative root mean-square errors (RMSE) of the detuning frequency, $[\mathbb{E}(\omega_{\rm noisy} - \omega)^2]^{1/2}/\omega$, as a function of $N_{\rm tot}$, for three different strategies: (i) the measurement of $X(t)$ for $N_{\rm times}=20$ equally spaced times; (ii) the measurement of $X(t)$ and  $Y(t)$ for $N_{\rm times}=20$ equally spaced times; (iii) the measurement of $X(t)$ at the two optimal times computed using the Fisher information; (iv) the measurement of $X(t)$ and $Y(t)$ at a single time. All the plots follow the expected $1/\sqrt{N_{\rm tot}}$ shot-noise dependence. The circles are the result of numerical simulations, and the dashed lines are the analytical results for the Cramér-Rao bound. Compared with the standard approach (strategy (i)), our strategy of measuring both quadratures at a single measurement time (strategy (iv)) reduces the square-root CRB bound by approximately $0.7$, corresponding to a $0.7^{2} \approx 50\%$ reduction in the number of shots required for a given precision. In addition, because $\langle Y(t)\rangle$ is antisymmetric with respect to $\omega$, strategy (iii) can determine both the amplitude \emph{and} the sign of $\omega$, while the other depends on the absolute value of $\omega$ only.

{\color{black}
The approach described so far suffers from circularity: the optimal times we computed require knowledge of $\omega$ and $\gamma$, the parameters we are trying to optimize. To address this problem, we assume that the detuning and decay rates change slowly over time, such that one can use their earlier values, $\omega_{\rm ref}$ and $\gamma_{\rm ref}$, to obtain an estimate, albeit not exact, of their current values. In this scenario, we need to check the resilience of the different methods to variation in the actual value of $\omega$ and $\gamma$ away from this reference point. This analysis is performed in Fig.~\ref{fig:single_qubit_noise_scaling}(b), where we fix the measurement times to their values optimized for $\omega_{\rm ref}=\gamma_{\rm ref}=1$ (i.e., $t_1\approx0.44/\gamma_{\rm ref}$ and $t_2\approx1.78/\gamma_{\rm ref}$ for the two-time strategy, and $t\approx1/\gamma_{\rm ref}$ for the joint quadrature strategy), and vary the actual dephasing rate $\gamma$ while keeping these times fixed. We again find that the fourth strategy (measuring $X$ and $Y$ at the same time) is the least sensitive to this mismatch between the reference and actual values of $\gamma$. A similar conclusion can be drawn for the stability to variations of $\omega$.
}

\section{Single qubit experiments}

\begin{figure*}[t]
    \centering
    \begin{tabular}{cc}
        \begin{overpic}[width=0.47\textwidth]{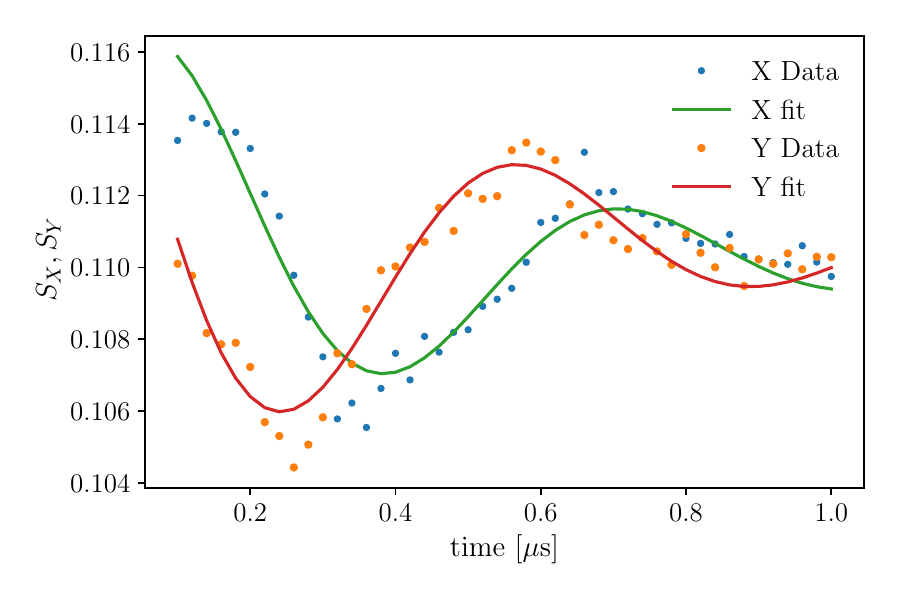}
            \put(-4,60){(a)}
        \end{overpic}
        &
        \begin{overpic}[width=0.47\textwidth]{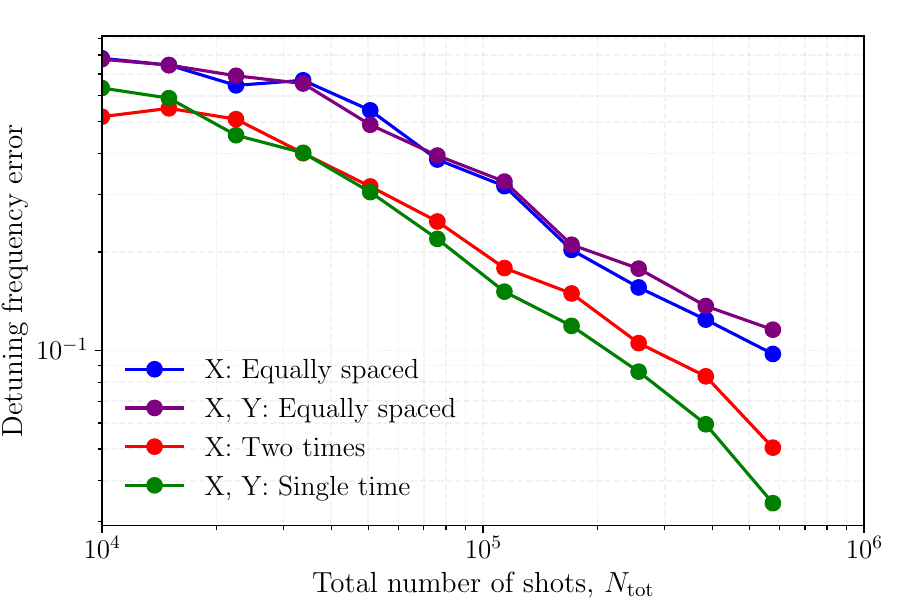}
            \put(-4,60){(b)}
        \end{overpic}
        \\[0.75em]
        \begin{overpic}[width=0.47\textwidth]{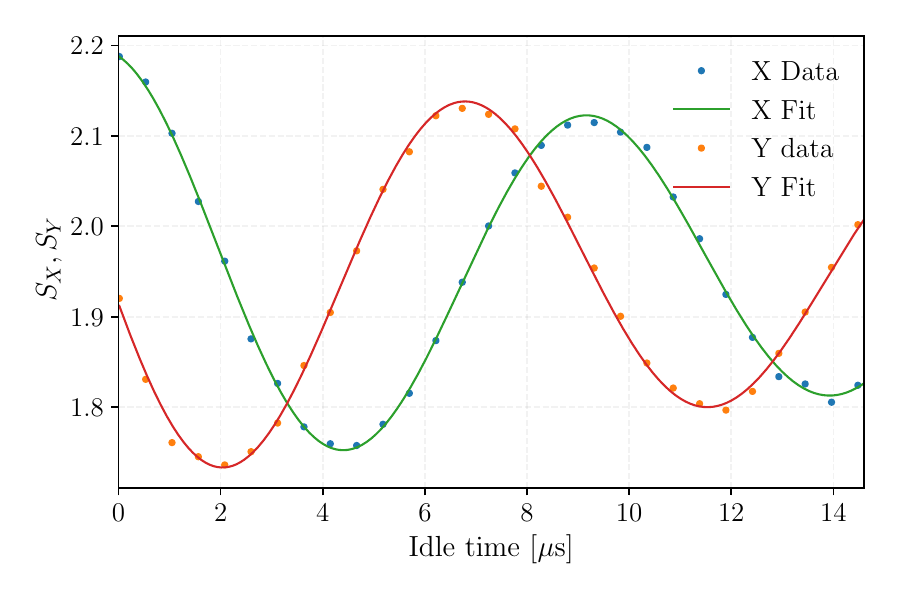}
            \put(-4,60){(c)}
        \end{overpic}
        &
        \begin{overpic}[width=0.47\textwidth]{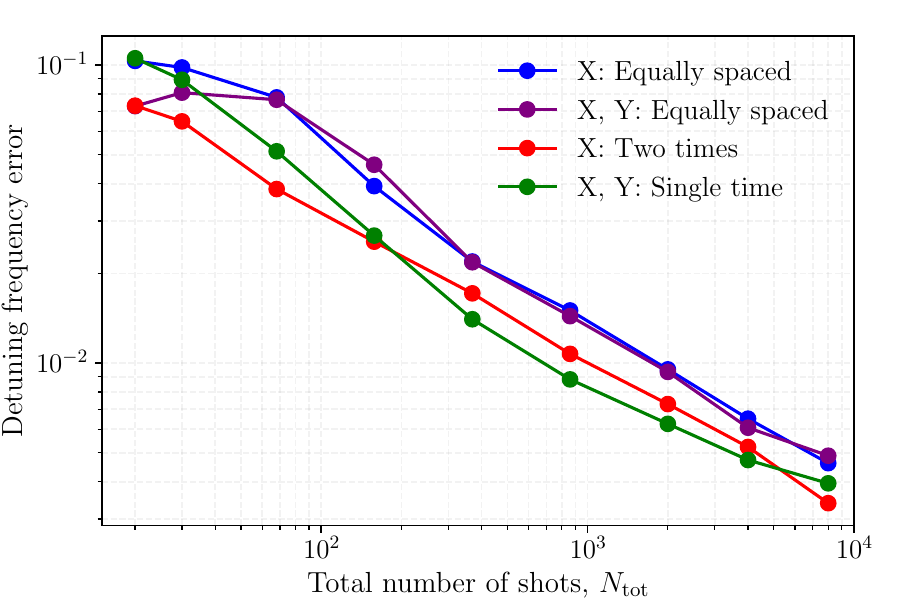}
            \put(-4,60){(d)}
        \end{overpic}
        
    \end{tabular}

    \caption{{\color{black}Experimental calibration of detuning and dephasing in quantum computing platforms.
            (a,b) Single NV-center experiment.
            (c,d) Single-transmon-qubit experiment.
            (a,c) Experimental Ramsey readout signals $S_X$ and $S_Y$ (dots) and their least-squares fits (solid lines), described by
            $
            S_X(t)=A_X\cos(\omega t+\phi_X)e^{-\gamma t}+B_X
            $
            and
            $
            S_Y(t)=-A_Y\sin(\omega t+\phi_Y)e^{-\gamma t}+B_Y,
            $
            as functions of time $\mathrm{\mu s}$.
            The signals are shown in experimental readout units rather than as normalized Pauli expectation values and are therefore not restricted to $[-1,1]$.
            The extracted fit parameters are used as reference values for the error estimation.
            (b,d) Error convergence versus the number of shots for different measurement strategies.
            In both platforms, measuring at the optimal times converges faster and systematically reduces the estimation error.}
            }
    \label{fig:single_qubit_results}
    \end{figure*}


{\color{black}
We conclude the analysis of the single-qubit case by presenting the results of two experiments, respectively using a single NV center and a superconducting transmon. 

The NV-center measurements were performed on an electronic-grade diamond sample containing etched nanopillars. Optical initialization and readout used a continuous-wave $520~\mathrm{nm}$ laser, and spin-dependent fluorescence around $630~\mathrm{nm}$ was detected using a single-photon counting module. A neodymium magnet provided a bias field of approximately $B_z\simeq 670~\mathrm{G}$, corresponding to a ground-state Zeeman splitting of about $1~\mathrm{GHz}$. Microwave control was applied through a bond-wire antenna, and the relevant $|0\rangle\leftrightarrow|1\rangle$ transition was identified by optically detected magnetic resonance, with
$
\omega_{\mathrm{NV}} = D-\gamma_e B_z,
\qquad
D=2.87~\mathrm{GHz},
\qquad
\gamma_e=2.8~\mathrm{MHz/G}.
$

The superconducting experiment was performed on Gilboa, a QuantWare Contralto-D chip consisting of 21 flux-tunable transmon qubits with fixed couplings and four readout lines. The average coherence times are $T_1,T_2 \sim 25~\mu\mathrm{s}$, and the qubit frequencies span approximately $4.9\text{--}6.6~\mathrm{GHz}$.

The experimental Ramsey data consisted of repeated measurements at $N_{\rm times}=46$ measurement times for the NV center and $N_{\rm times}=30$ measurement times for the transmon qubit, with up to $N_{\rm shots}=10^5$ and $N_{\rm shots}=10^3$ shots per time point, respectively.\footnote{Note that the experimental measurements of the NV center are of a statistical nature as (i) they rely on a weak spin-dependent fluorescence difference of about 30\% and (ii) the collection efficiency of photons is approximately 10\%. These two effects explain the difference of two orders of magnitude in the number of shots used in the two experiments.}

{\color{black}
For each system, we first used the full experimental datasets to obtain reference, or ``ground-truth'', fit parameters for the two fitting procedures considered below. The experimental readout signal $S_X(t)$ is related to the normalized Ramsey response $\langle{X(t)}\rangle$ by an amplitude scaling and an offset that account for the experimental readout calibration. For the X-only strategies, the full X-quadrature dataset was therefore fitted to

\begin{equation}
    S_X(t)=A_X \langle{X(t)}\rangle+B_X,
\end{equation}

where

\begin{equation}
\langle{X(t)}\rangle=\cos(\omega_Xt+\phi_X)e^{-\gamma_Xt}.
\end{equation}

}
This fit provided the reference values used when evaluating the reduced-data $X$-only estimators. Separately, for the joint-quadrature strategy, the full $X$ and $Y$ datasets were fitted simultaneously using the shared parameters $\omega_{XY}$ and $\gamma_{XY}$, with quadrature-dependent amplitudes, phases, and offsets. These joint-fit parameters provided the reference values for the reduced-data $X/Y$ estimator. In this way, each reduced-data strategy was compared to the corresponding high-shot fit obtained using the same measurement model. In the reduced-data fits, the nuisance parameters $A$, $\phi$, and $B$ are fixed to their reference values, so that only $\omega$ and $\gamma$ are treated as free parameters, consistent with the two-parameter Fisher-information optimization derived above. See table \ref{tab:reference_fit_parameters} for the calculated reference values.

\begin{table*}[t]
\centering
\caption{Reference fit parameters extracted from the full Ramsey datasets. 
The angular frequency $\omega$ is reported in $\mathrm{rad}/\mu\mathrm{s}$, 
and the decay rate $\gamma$ is reported in $\mu\mathrm{s}^{-1}$. 
The amplitudes and offsets are given in the native, dimensionless readout units.}
\label{tab:reference_fit_parameters}
\begin{tabular}{llcccccc}
\hline
System & Fit & $A$ & $\phi$ [rad] & $B$ 
& $\omega$ [$\mathrm{rad}/\mu\mathrm{s}$] 
& $\gamma$ [$\mu\mathrm{s}^{-1}$] 
& $t_{\rm opt}$ [$\mu\mathrm{s}$] \\
\hline
NV & $X$
& $0.006$ & $-0.845$ & $0.110$ & $9.782$ & $1.696$ 
& $0.456,\ 0.647$ \\

NV & $X/Y$  
& $-0.007$ & $2.343$ & $0.110$ & $9.710$ & $2.181$ 
& $0.420$ \\

Transmon & $X$
& $0.238$ & $0.192$ & $1.954$ & $0.659$ & $0.037$ 
& $13.448,\ 15.0$ \\

Transmon & $X/Y$  
& $0.238$ & $0.185$ & $1.954$ & $0.660$ & $0.037$ 
& $14.48$ \\
\hline
\end{tabular}
\end{table*}

To compare different calibration strategies on the same experimental dataset, we performed a random downsampling analysis. For each total shot budget $N_{\rm tot}$, a subset of shots was drawn without replacement from the full experimental record and assigned according to the measurement allocation prescribed by each strategy. We considered: (i) uniform $X$ quadrature sampling over all available measurement times, (ii) $X$ quadrature measurements concentrated at the two Fisher-optimized measurement times, (iii) joint $X$ and $Y$ quadrature measurements at a single optimized measurement time, and (iv) uniform joint $X$ and $Y$ quadrature sampling over all available measurement times. For each strategy, the available $N_{\rm tot}$ shots were equally distributed between the measurement times and quadratures.
This random selection procedure was repeated 300 times in order to obtain a high-precision estimate of the average estimation error for each strategy.


The optimized measurement times were chosen by minimizing the Cramér--Rao bound computed from the corresponding reference parameters. This choice emulates the practical situation in which a previous calibration provides an estimate of the relevant timescale, which is then used to design a faster subsequent calibration. For the two-time $X$ strategy, the Fisher information was evaluated using the full $X$-fit values of $\omega_X$, $\gamma_X$, and $\phi_X$. The optimized joint $X/Y$ strategy was evaluated using the corresponding parameters obtained from the full joint fit. In both cases, the cost function used the Cramér--Rao variances for $\omega$ and $\gamma$, normalized by their respective reference values, so that the optimization balanced the relative variances of the two parameters. The phase $\phi$ entered the Fisher-information calculation through the Ramsey model and therefore influenced the optimal measurement times. The resulting optimized times were then used in the downsampling analysis described above.

For the transmon joint-quadrature strategy, the Fisher optimization predicted an optimal measurement time beyond the experimentally sampled Ramsey trace. Because the transmon data contained a relatively small number of measurement times, $N_{\rm times}=30$, and the theoretical optimum for the $X/Y$ method lay outside the available time window, we did not optimize over continuous measurement times for both $X$ and $X/Y$ cases. Instead, we restricted the optimization to the discrete set of experimentally available measurement times and selected the best time within that set. This discretization was applied only to the experimental downsampling analysis and does not change the Fisher information criterion itself.

The results of this analysis for the NV center are shown in Fig.~\ref{fig:single_qubit_results} (a,b) and for the superconducting transmon in Fig.~\ref{fig:single_qubit_results} (c,d). As expected, all four methods improve as the total number of shots is increased. The main difference is how efficiently each method uses the available shots. For the $X$ only data, concentrating the measurements at the two optimized times gives a clear improvement over spreading the same number of shots uniformly over all measurement times.

The same trend is seen for the joint X/Y strategies. In the NV-center data, the optimized single-time X/Y measurement gives the lowest detuning-frequency error over most of the range shown. In the transmon data, the optimized joint X/Y strategy also outperforms the optimized X-only strategy, despite the fact that the Fisher-optimal X/Y measurement time lies beyond the experimentally sampled time window and was therefore constrained to the available measurement times. 

{\color{black}
\begin{table}[t]
\centering
\small
\setlength{\tabcolsep}{4pt}
\caption{
Percentage reduction in the number of shots required to reach the same
detuning-estimation error as uniform $X$-quadrature sampling.
}
\label{tab:single_qubit_shot_reduction}
\begin{tabular}{lccc}
\hline
\rule{0pt}{4.8ex}

System 
& \shortstack{$X$: \\two times}
& \shortstack{$X,Y$: \\single time}
& \shortstack{$X,Y$: \\ equally spaced} \\
\hline
NV center 
& 50.4\% 
& 57.6\% 
& $-7.2\%$ \\

Transmon 
& 49.4\% 
& 49.0\% 
& 10.4\% \\
\hline
\end{tabular}
\end{table}

To quantify the reduction in measurement resources, for each reported error value of a given strategy we estimated how many shots would be required by the
uniformly sampled $X$ strategy to reach the same error. This was done by linearly interpolating between neighboring points of the equally-spaced- $X$ curve in log-log space (Fig.~\ref{fig:single_qubit_results} b,d). The ratio between this interpolated shot count and the number of shots used by the tested strategy gave the shot-reduction factor at that error, for which we then calculated the geometric mean over all possible points and converted it to a percentage reduction in the required number of shots. The results are summarized in table ~\ref{tab:single_qubit_shot_reduction}.
 For the NV-center experiment, the two-times $X$ and single-time $X, Y$ strategies reduce the required number of shots by approximately $ 50\%$ and $ 58\%$, respectively. For the single-transmon experiment, both optimized strategies yield a reduction of approximately $49\%$.

}
Both optimized strategies give a clear improvement over their uniformly sampled counterparts. This is consistent with the Fisher-information picture: concentrating shots near the most informative accessible measurement times avoids the loss of precision associated with spending shots at less informative points in the Ramsey trace.

Overall, the experimental data support the main prediction of the theory. For a fixed shot budget, the choice of measurement times matters. Using the Cramér--Rao bound to select a small number of informative measurement times can reduce the number of shots needed to reach a given frequency precision, compared with uniform sampling over the full time trace.

}

\section{Two-qubit results}


{\color{black}
We now move to coupled-qubit systems relevant to quantum processors. 
Residual \(ZZ\) interactions are a prominent coherent crosstalk mechanism in 
superconducting devices, where they appear as shifts of a qubit's transition 
frequency conditioned on the computational state of neighboring qubits.
Such conditional shifts have been experimentally observed and mitigated in 
several architectures \cite{mundada2019suppression,ni2022scalable,xie2022suppressing}, and their microscopic origin has been analyzed in 
recent theoretical works 
\cite{heng2024estimatingeffectcrosstalkerror,
fors2024comprehensiveexplanationzzcoupling}. 
We therefore model the coupled system by local detunings together with 
nearest-neighbor static \(ZZ\) terms. 

Let us first consider the case of two qubits
\begin{align}
\label{eq:Htwo}
\begin{split}
H = \sum_{i=1,2} 
    \left[\tfrac{\omega_i}{2} + \tfrac{h_i(t)}{2}\right](1 - Z_i)
    \\
    + \tfrac{J}{4}(1 - Z_1)(1 - Z_{2}).
\end{split}
\end{align}

Here, $\omega_i$ is the unknown detuning of qubit (i), defined in the rotating frame used for the Ramsey experiment, and $(h_i(t))$ is a stochastic dephasing field with zero mean, as in the single-qubit model. The parameter J describes a static nearest-neighbor ZZ coupling. Because the interaction term is proportional to $((1-Z_1)(1-Z_2))$, it contributes only when both qubits are in the $\ket{1}$ state. Consequently, the Ramsey frequency of qubit 1 is $\omega_1$ when qubit 2 is prepared in $\ket{0}$, and $\omega_1 + J$ when qubit 2 is prepared in $\ket{1}$. Equivalently, the Ramsey frequency of qubit 2 is shifted by J depending on the state of qubit 1. Thus, J can be extracted from neighbor-state-dependent Ramsey frequency shifts. Since Eq.~(\ref{eq:Htwo}) is diagonal in the computational basis, its dynamics can be solved analytically for any initial state.

{\color{black}

The protocol, therefore, consists of two Ramsey measurements on the first qubit. In the first, the neighboring qubit is prepared in $\ket{0}$, so that the measured Ramsey frequency gives the bare detuning of the first qubit. In the second, the neighboring qubit is prepared in $\ket{1}$, and the measured Ramsey frequency includes the conditional shift induced by the ZZ interaction. The crosstalk coupling is then obtained as the difference between these two estimated frequencies.

Similar Ramsey-based approaches for measuring crosstalk have been proposed and demonstrated previously \cite{li2024realization, jurcevic2022effective}. Our goal here is not to introduce a fundamentally different crosstalk observable, but rather to formulate the calibration problem in a way that naturally fits the optimized single-qubit Ramsey framework developed above. We therefore separate the estimation of bare detunings from the estimation of conditional frequency shifts: the first pair of experiments determines the detunings, $\omega_i$, while the second pair measures the shifted frequencies $\omega_i+J_{i\pm1}$, from which the crosstalk couplings are extracted. With this organization, each stage can be treated using the same Fisher-information analysis as a single-qubit detuning-estimation problem. Other estimation algorithms, including JAZZ-type protocols \cite{takita2017experimental}, could be applied to the same physical problem and optimized using Fisher information; the distinction here is that we explicitly organize the calibration into separate detuning and crosstalk stages.

More generally, our approach should be viewed as complementary to crosstalk mitigation techniques rather than as an alternative to them. Several works have proposed methods to suppress, compensate, or dynamically decouple unwanted ZZ interactions \cite{tripathi2022suppression, mundada2019suppression, ni2022scalable, xie2022suppressing}. These strategies rely on characterizing the conditional frequency shifts produced by crosstalk, either to choose compensation parameters or to benchmark the effectiveness of the mitigation. The contribution of the present work is to optimize this parameter-estimation step: for a fixed measurement budget, we identify Ramsey times and quadratures that minimize the statistical uncertainty in the extracted detunings and crosstalk couplings. Thus, the optimized measurement protocol developed here can serve as input to subsequent crosstalk-compensation or suppression procedures.}

We demonstrated the feasibility of this approach by calibrating the crosstalk coupling between two transmon qubits in the Gilboa superconducting quantum computer \cite{iqcc}, with up to $N_{\rm times}=30$ and $N_{\rm shots} = 20000$.

\begin{figure}[t]
	\centering
	\begin{tabular}{l} 
    \begin{overpic}[width=0.95\linewidth]{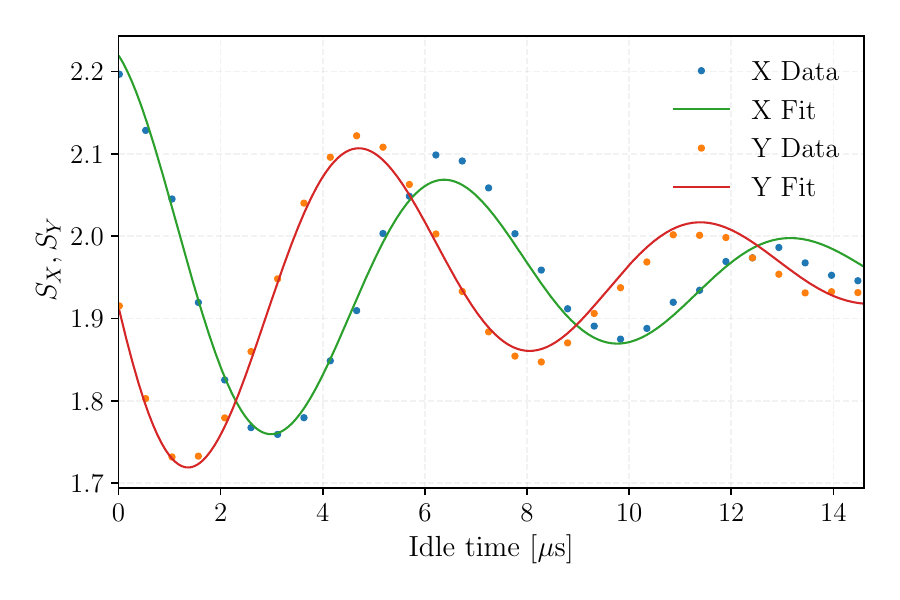}
    \put(0,60){(a)}
  \end{overpic}\\
    \begin{overpic}[width=0.95\linewidth]{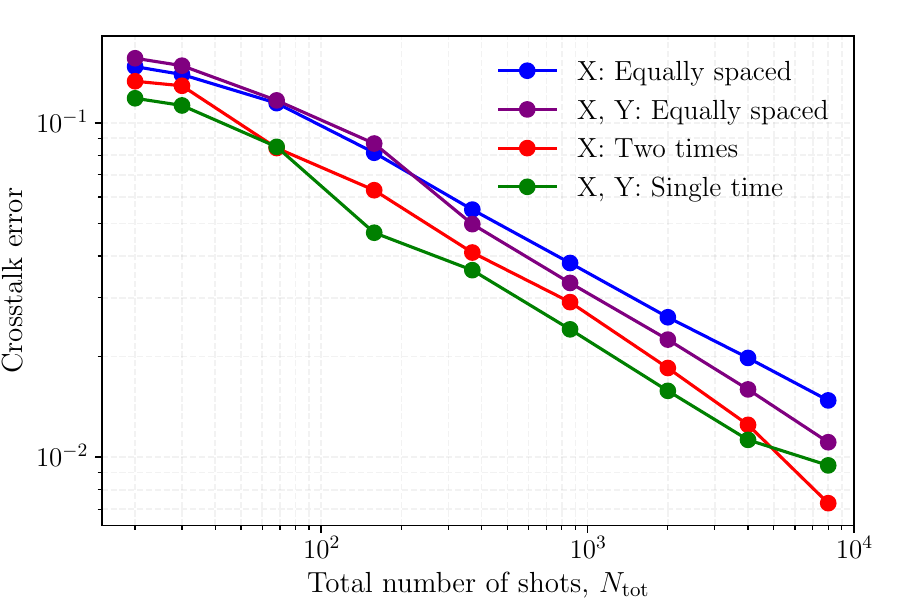}
    \put(0,60){(b)}
  \end{overpic}
	\end{tabular}
    \caption{{\color{black}Experimental calibration of detuning, dephasing, and crosstalk in a superconducting multi-qubit system. (a) Experimental Ramsey readout signals $S_X$ and $S_Y$ (dots) and their least-squares fits (solid lines), described by
        $S_X(t)=A_X\cos(\omega t+\phi_X)e^{-\gamma t}+B_X$        
        and        
        $S_Y(t)=-A_Y\sin(\omega t+\phi_Y)e^{-\gamma t}+B_Y$,        
        as functions of time $\mathrm{\mu s}$, for a target qubit whose nearest neighbor is prepared in $\ket{1}$. The signals are shown in experimental readout units rather than as normalized Pauli expectation values and are therefore not restricted to $[-1,1]$. (b) Comparison of four different strategies to calibrate the crosstalk between two superconducting qubits. The same two-quadrature protocol again shows the best scaling with the number of shots.}}
    \label{fig:exp2}
\end{figure}

We followed an analogous procedure to characterize the crosstalk coupling between two transmon qubits. As in the single-qubit case, we first performed a Ramsey interference experiment on the target qubit with its neighbor prepared in the $\ket{0}$ state, extracting a detuning $\omega_0$. We then repeated the same measurement with the neighboring qubit prepared in the $\ket{1}$ state, extracting a shifted detuning $\omega_1$. The crosstalk coupling is given by the difference between the two, $J = \omega_1 - \omega_0$. The raw Ramsey traces for for a target qubit whose nearest neighbor is prepared in $\ket{1}$ are shown in Fig.~\ref{fig:exp2}(a).

For each neighbor preparation, we obtained reference, or ``ground-truth,'' fit parameters from the full experimental dataset, following the same procedure described above for the single-qubit case: the full $X$ quadrature data were fitted to the $X$ only model to obtain reference values for the $X$ only strategies, while the full $X$ and $Y$ datasets were jointly fitted to obtain reference values for the joint-quadrature strategy. The reference crosstalk coupling $J$ was then obtained from the difference between the reference detunings extracted for the two neighbor preparations. See table \ref{tab:transmon_reference_fit_parameters} for the calculated reference values.

\begin{table*}[t]
\centering
\caption{Reference fit parameters and optimized measurement times for the transmon Ramsey dataset.}
\label{tab:transmon_reference_fit_parameters}
\begin{tabular}{lcccccc}
\hline
Fit & $A$ & $\phi$ [rad] & $B$ 
& $\omega$ [$\mathrm{rad}/\mu\mathrm{s}$] 
& $\gamma$ [$\mu\mathrm{s}^{-1}$] 
& $t_{\rm opt}$ [$\mu\mathrm{s}$] \\
\hline
$X$ 
& $0.279$ & $0.246$ & $1.947$ & $0.926$ & $0.129$ & $5.696,\ 7.248$ \\

joint $X/Y$ 
& $0.279$ & $0.224$ & $1.947$ & $0.933$ & $0.131$ & $7.248$ \\

\hline
\end{tabular}
\end{table*}

We compared the same four measurement strategies considered in the single-qubit analysis: uniform $X$-quadrature sampling, $X$ quadrature sampling concentrated at the two Fisher-optimized times, joint $X/Y$ sampling at a single optimized time, and uniform joint $X/Y$ sampling. For each strategy and each total shot budget $N_{\rm tot}$, the available shots were randomly downsampled from the full experimental record, separately for each neighbor preparation, and the resulting fits were used to extract $J$. Here, $N_{tot}$ denotes the number of shots per neighbor preparation, so the total number of physical shots used to estimate J is $2N_{tot}$. This procedure was repeated 300 times per shot budget to obtain a high-precision estimate of the average estimation error for each strategy.

The results of this analysis are shown in Fig.~\ref{fig:exp2}(b). We find that, over almost the entire range of shot budgets considered, the relative performance of the four strategies closely mirrors that observed in the single-qubit case: 
{\color{black}
The optimized two-time $X$-only strategy reduces the required number of shots by an average of about $49\%$ relative to uniform $X$ sampling. The single-time $X/Y$ strategy gives a $64\%$ reduction, while uniform $X/Y$ sampling gives a $15\%$ reduction.
}
This indicates that the advantage of concentrating measurements at Fisher-optimized times and quadratures, established for single-qubit detuning estimation, extends directly to the estimation of crosstalk couplings in coupled-qubit systems.


\section{Multi-qubit extension}
We now move to the case of many qubits, but considering a 
one-dimensional chain described by
\begin{align}
\label{eq:Hmany}
\begin{split}
H = \sum_{i=1}^{N_{\rm qubits}} 
\Big(
    \left[\tfrac{\omega_i}{2} + \tfrac{h_i(t)}{2}\right](1 - Z_i)
    \\
    + \tfrac{J_i}{4}(1 - Z_i)(1 - Z_{i+1})
\Big).
\end{split}
\end{align}

A naive approach to estimate all system parameters ($\vec{\omega}$, $\vec{\gamma}$, $\vec{J}$) consists of initializing all qubits in the $|+\rangle$ state, allowing them to evolve over time, and subsequently measuring all qubits simultaneously. The resulting data can then be fitted to the analytical model to deduce the system's parameters, including both individual qubit detunings and crosstalk effects. 

Although the physical observables depend on all system's parameters, extracting them is not as efficient as in the single-qubit case. Due to the finite number of shots, we incur inevitable uncertainty in our data, governed by $\frac{1}{\sqrt{N}}$. This creates the problem of overfitting the noisy data to a different set of parameters. We empirically observe that using this method, the estimation errors are approximately an order of magnitude higher than those obtained from the single-qubit methods.

An alternative strategy for computing the values of $\omega_i$, $\gamma_i$, and $J_i$ involves reducing the many-body problem to that of isolated single qubits. Specifically, we propose to perform two pairs of experiments, respectively, to probe the detunings and the crosstalk couplings. see Fig.~\ref{fig:chain} for details. Each experiment involves a Ramsey interference experiment on half of the qubits (denoted by ``H''), for different initial states of the other qubits. In the former two experiments, the qubits rotate at frequency $\omega_i$, while in the latter two, they rotate at frequency $\omega_i + J_{i\pm 1}$. 
}

%
%


\begin{figure}[t]
	\centering
    \begin{tabular}{c}
    (a)\\
	\includegraphics[width=0.9\linewidth]{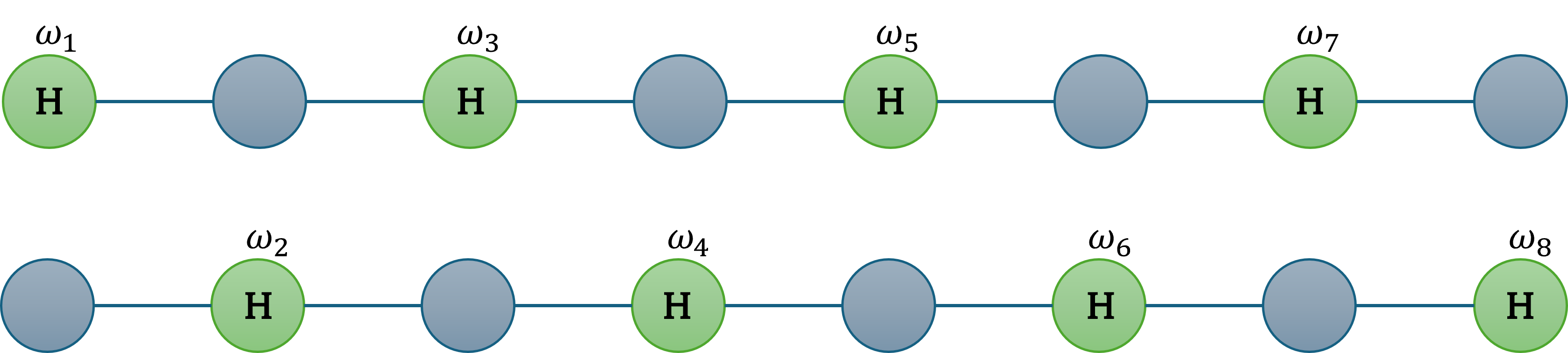}\\
    (b)\\\includegraphics[width=0.9\linewidth]{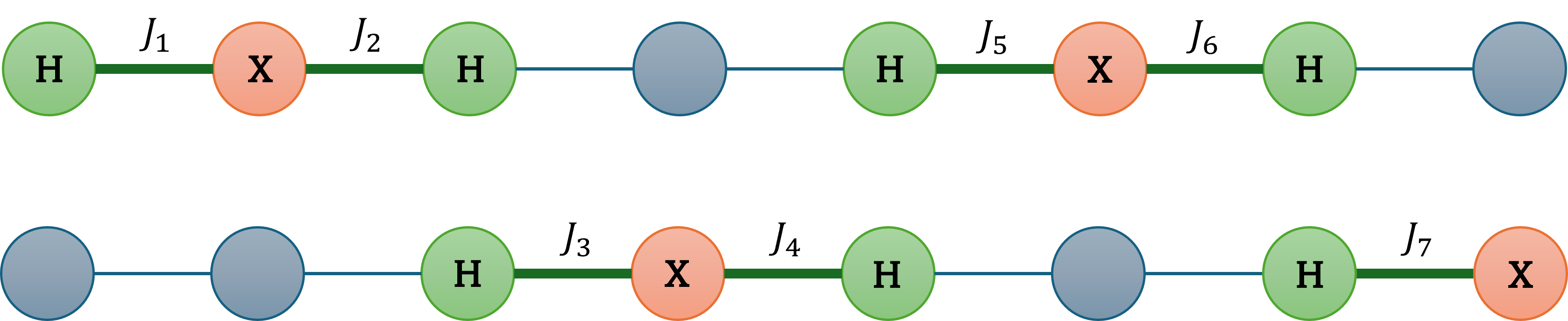}
    \end{tabular}
	\caption{Schematic representation of the four experiments used to calibrate the multi-qubit system defined in Eq.~(\ref{eq:Hmany}): (a) two experiments are used to measure $\omega_j$: the green nodes represent qubits in the $|+\rangle$ state, and the blue nodes represent qubits in the $|0\rangle$ state; (b) two experiments are used to measure $J_j$: the red nodes represent qubits in the $\ket{1}$, the green nodes represent qubits in the $\ket{+}$ state, and the blue nodes represent qubits in the $\ket{0}$ state. The green edges are the measured crosstalk.}
	\label{fig:chain}
\end{figure}


{\color{black} This approach can be straightforwardly extended to quantum computers with more complex qubit connectivity.} In general, the number of experiments needed to calibrate the system depends on the amount of non-negligible crosstalk couplings per qubit. For bipartite lattices with nearest-neighbor couplings, it is possible to see that one can determine all the systems' parameters using the same four experiments as in the one-dimensional case. Interestingly, for IBM's heavy-hex topology, it can be shown that four experiments are also sufficient. We demonstrate this extension to the two-dimensional topology of IBM quantum computers in Fig.~\ref{fig:IBM-tiling}. The figure illustrates the tiling required for probing crosstalk effects, where red nodes represent qubits in the $\ket{+}$ state, where the Ramsey interference is performed, and blue (yellow) nodes represent qubits in the $\ket{0}$ ($\ket{1}$) state.

{\color{black} The tiling in Fig.~\ref{fig:IBM-tiling} is obtained by applying the same isolation principle used in the one-dimensional chain. Qubits on which Ramsey oscillations are measured are chosen so that no two of them are nearest neighbors. Otherwise, the frequency shift of a measured qubit would depend on the evolving state of another measured qubit. The remaining qubits are initialized in computational basis states. To measure many crosstalk terms in a single experiment, the qubits initialized in $\ket{1}$ are chosen to be adjacent to as many measured qubits as possible. Each neighboring measured qubit then acquires a conditional frequency shift from that $\ket{1}$ qubit. A second complementary tiling measures the remaining edges. Together with the two detuning experiments, this gives a four-experiment protocol for the heavy hex geometry, the same number required for the one-dimensional chain.}

\begin{figure}[t]
    \centering
    \begin{minipage}[b]{0.48\linewidth}
        \centering
        \includegraphics[width=\linewidth]{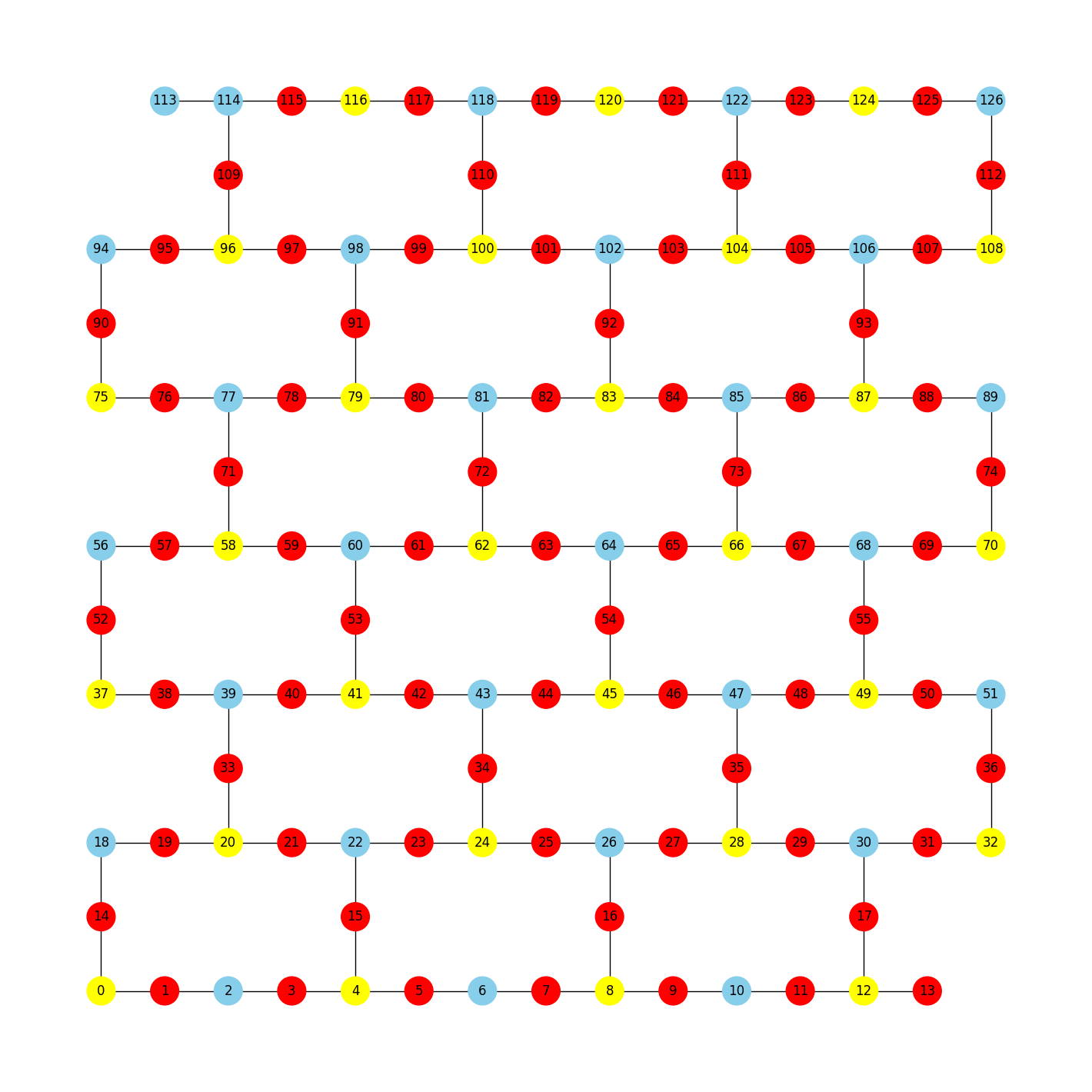}
    \end{minipage}
    \hfill
    \begin{minipage}[b]{0.48\linewidth}
        \centering
        \includegraphics[width=\linewidth]{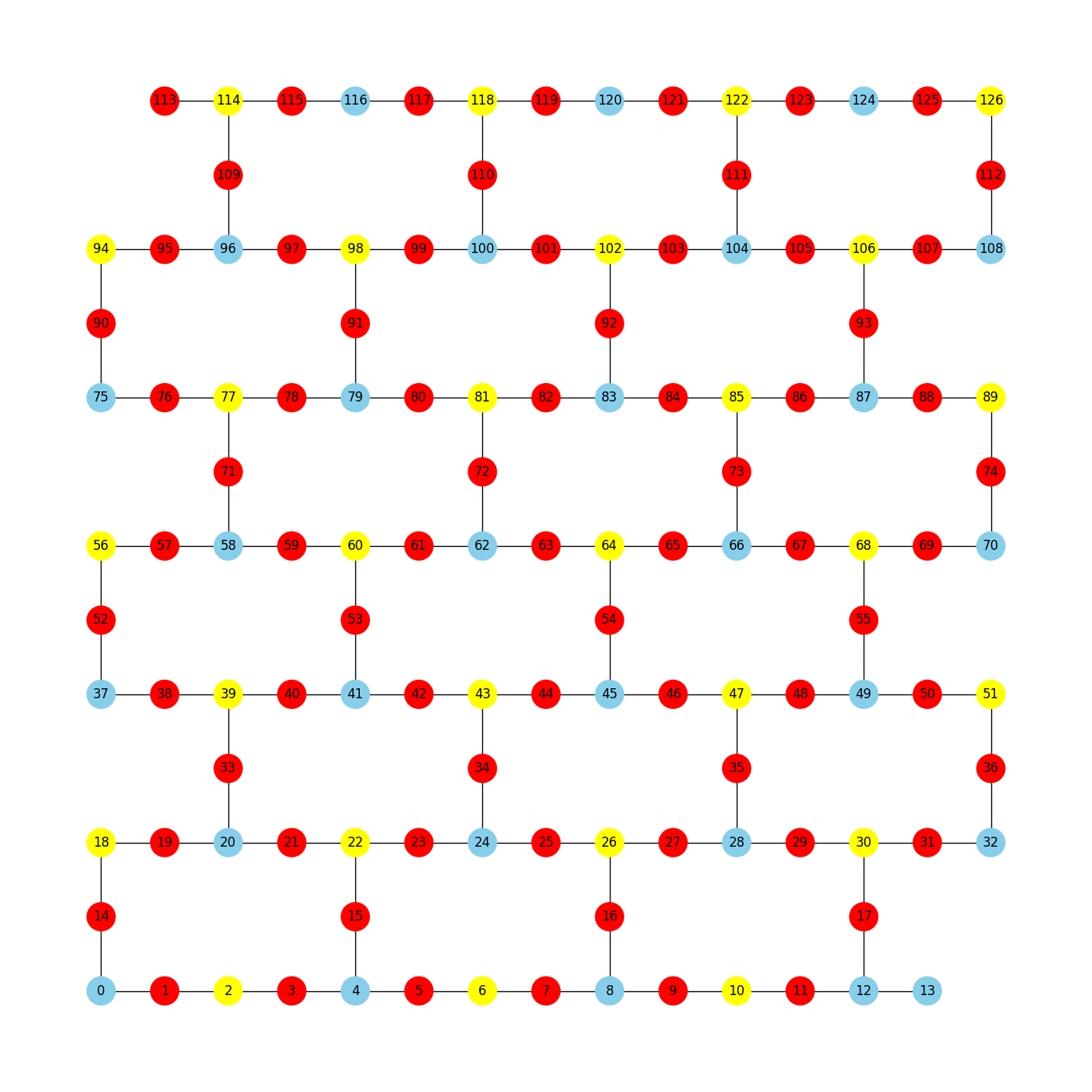}
    \end{minipage}

    \caption{IBM tiling for probing crosstalk effects. The red nodes represent qubits in the $\ket{+}$ state where the Ramsey interference is performed, and blue (yellow) nodes represent qubits in the $\ket{0}$ ($\ket{1}$) state.}
    \label{fig:IBM-tiling}
\end{figure}

For more complex topologies, the problem at hand can be formulated as a tiling optimization problem, which can be solved heuristically and warrants further investigation.

\section{Summary and discussion}

This article proposes and validates optimal strategies for characterizing qubit detuning and crosstalk in superconducting quantum computers. While traditional methods spread measurements across multiple times to average out noise, our work demonstrates that concentrated, optimally-timed measurements can achieve better results with fewer resources. By analyzing different measurement protocols using the Fisher information framework and the Cramér–Rao bound, we demonstrated that measurements of both X and Y quadratures at a single optimal time $t\approx 1/\gamma$ yields the best precision and resilience to parameter variations. 
We then extended our analysis to multi-qubit systems, where we described a method that effectively reduces the estimation of crosstalk terms to decoupled single-qubit problems, drastically simplifying calibration. Our framework scales well to larger systems and complex architectures, making it highly relevant for current and future quantum processors. Experimental validation and simulations confirm that the proposed strategies can cut calibration time by up to 50\% without compromising accuracy. Looking forward, integrating these optimized calibration strategies into real-time control systems could significantly enhance the stability and scalability of quantum computing platforms.



\acknowledgments{We acknowledge useful discussions with N. Bar-Gill, S. Burov, and O. Hamdi.  We are thankful to N. Alfasi and O. Ovdat from the Israeli Quantum Computing Center (IQCC) for technical support. D.S. and E.G.D.T. are supported by the Israel Science Foundation, grants 2126/24 and 2471/24. }
        
\section*{Author Contributions}

All authors contributed equally. Large language models were
used only for bibliographic searches and language editing.

\bibliographystyle{quantum}
\bibliography{main}


\onecolumngrid
\appendix

\end{document}